\documentclass[pre,twocolumn,floatfix,aps,showpacs]{revtex4}
\usepackage{amssymb}
\usepackage{graphicx}
\usepackage{epsf}
\usepackage{amsmath}
\usepackage{bm} 
\usepackage{pspicture}
\usepackage{flafter}
\usepackage{float}
\begin{document}

\title{Effects of Disorder on Synchronization of Discrete Phase-Coupled
Oscillators}
\author{Kevin Wood$^{1,2}$, C. Van den Broeck$^{3}$, R. Kawai$^{4}$,
and Katja Lindenberg$^{1}$}
\affiliation{
$^{(1)}$Department of Chemistry and Biochemistry and Institute for
Nonlinear Science, and $^{(2)}$ Department of Physics,
University of California San Diego, 9500 Gilman Drive, 
La Jolla, CA 92093-0340, USA\\
$^{(3)}$Hasselt University, Diepenbeek, B-3590 Belgium\\
$^{(4)}$ Department of Physics, University of Alabama at Birmingham,
Birmingham, AL 35294 USA
}
\date{\today}

\begin{abstract}
We study synchronization in populations of phase-coupled stochastic
three-state oscillators characterized by a distribution of
transition rates.  We present results on an exactly solvable
dimer as well as a systematic characterization of globally connected
arrays of $\mathcal{N}$ types of oscillators ($\mathcal{N}=2$, $3$,
$4$) by exploring the linear stability of the nonsynchronous fixed
point.  We also provide results for globally
coupled arrays where the transition rate of each unit is drawn
from a uniform distribution of finite width.  Even in the presence
of transition rate disorder, numerical and analytical results
point to \emph{a single phase transition} to macroscopic synchrony at a
critical value of the coupling strength. 
Numerical simulations make possible the further characterization
of the synchronized arrays. 
\end{abstract}
\pacs{64.60.Ht, 05.45.Xt, 89.75.-k}

\maketitle

\section{Introduction}
Synchronization in populations of phase-coupled nonlinear stochastic
oscillators, and the corresponding emergence of macroscopic coherence,
appear pervasively in a tremendous range of physical, chemical, and
biological systems. As a result, the general subject
continues to be studied intensely in applied mathematics and theoretical
physics~\cite{strogatz,winfree,kuramoto,strogatz2,pikovsky}. 
Since the pioneering
work of Kuramoto~\cite{kuramoto}, emergent cooperation in these systems
has been investigated from a myriad of perspectives encompassing both
globally and locally coupled, stochastic and deterministic, and large
and small systems.  And while Kuramoto's canonical model of
nonlinear oscillators, whose use is widespread because of its close
kinship to the normal form describing general phase oscillations, has
proven spectacularly successful for characterizing synchronization,
simple, phenomenological models of synchronization have also proven
useful in a variety of new
contexts~\cite{lutz,local,threestate1,threestate2},
most notably the characterization of emergent synchronization as
a nonequilibrium phase transition. 

We have shown~\cite{threestate1,threestate2}
that a model of three-state identical phase-coupled stochastic
oscillators is ideally suited for studying the nonequilibrium phase
transition to synchrony in locally-coupled systems,
owing in large part to its numerical simplicity. 
The utility of these studies rests on the well-established notion
of universality, that is, on the contention that microscopic details
do not determine the universal properties associated with the
breaking of time-translational symmetry that leads to a macroscopic
phase transition. 
Statistical mechanics is thus enriched by simplistic, phenomenological
models (the Ising model being the most ubiquitous example) whose
microscopic specifics are known to be, at best, substantial simplifications
of the underlying quantum mechanical nature of matter, but whose
critical behavior captures that of more complex real systems.
In this spirit, our simple tractable model captures the principal
features of the syncrhonization of phase-coupled oscillators. 
In the globally coupled (mean field) case our model undergoes a
supercritical Hopf bifurcation. With nearest neighbor coupling,
we have shown that the array undergoes a continuous
phase transition to macroscopic synchronization marked by signatures of the
$XY$ universality class~\cite{risler2,risler}, including the appropriate
classical exponents $\beta$ and $\nu$, and lower and upper critical
dimensions $2$ and $4$ respectively.  

In this paper we focus on globally coupled arrays and expand our earlier
studies to the arena of transition rate disorder. We start with a slightly
modified version of our original model (explained below), in which
identical synchronized units are governed by the same transition rates
as individual uncoupled units.  Then, in the spirit of
the original Kuramoto problem~\cite{kuramoto}, we explore the occurrence
of synchronization when there is more than one transition rate and perhaps
even a distribution of transition rates among the phase-coupled
oscillators. In particular, we explore the conditions (if any) that lead
to a synchronization transition in the face of a transition rate distribution,
discuss the relation between the frequency of oscillation of the
synchronized array and the transition rates of individual units, and
explore whether or not the existence of units of different transition
rates in the coupled array may lead to more than one phase transition.

Our paper is organized as follows. In Sec.~\ref{sec2} we describe the
model, including the modification (and associated rationale) 
of our earlier scenario even for an array of identical
units.  In Sec.~\ref{sec3} we present results for
a dimer composed of two units of different intrinsic transition rates.  While
there is of course no phase transition in this system, it is instructive
to note that there is a probability of sychronization of the two units
that increases with increasing coupling strength.
Section~\ref{sec4} introduces disorder of a
particular kind, useful for a number of reasons that include some
analytical tractability.  Here our oscillators can have only one of
$\mathcal{N}$ distinct transition rates, where $\mathcal{N}$ is a small
number.  We pay particular attention to the dichotomous case,
$\mathcal{N}=2$. This simple disordered system reveals
some important general signatures of synchronization. We also consider
the cases $\mathcal{N}=3$ and $\mathcal{N}=4$, but find that the
$\mathcal{N}=2$ case already exhibits most of the 
interesting qualitative consequences of
a distribution of transition rates.  In particular, we are
able to infer the important roles of
the mean and variance of the distribution.
In Sec.~\ref{sec5} we generalize further to a uniform finite-width
distribution of transition rates and explore this inference in more detail.
Section~\ref{sec6} summarizes our results and poses some questions for
further study.

\section{The Model}
\label{sec2}
\begin{figure}[b]
\begin{center}
\includegraphics[width=3.0 cm]{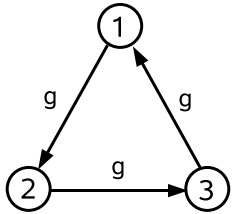}
\caption{Three-state unit with transition rates $g$.}
\label{fig1}
\end{center}
\end{figure}

Our point of departure is a stochastic three-state model governed by
transition rates $g$ (see Fig.~\ref{fig1}), where each state may be
interpreted as a discrete phase~\cite{threestate1,threestate2}. 
The unidirectional, probabilistic nature of the transitions among
states assures a qualitative analogy between this three-state discrete
phase model and a noisy phase oscillator.  
\begin{figure}
\includegraphics[width=7 cm]{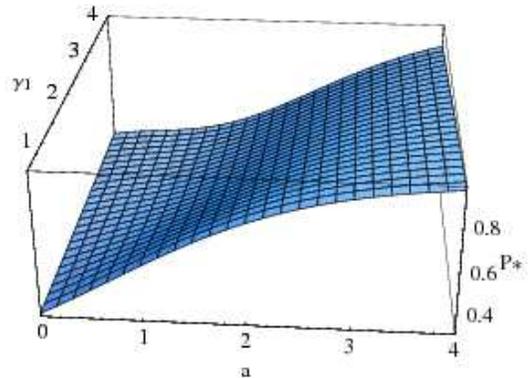}
\caption{(Color online) Phase space surface of the steady state probability
$\mathcal{P}_A^*$ that both units
of a dimer are in the same state, indicating perfect
synchronization, for $\gamma_2=0.5$ and a range of $\gamma_1$ and $a$. 
}
\label{dimer1}
\end{figure}
The linear evolution equation of a single oscillator is
$\partial P(t)/\partial t = M P(t)$, where the components $P_i(t)$ of the
column vector $P(t)= (P_1(t)~P_2(t)~P_3(t))^T$ ($T$ denotes the
transpose)
are the probabilities of being in states $i=1,2,3$ at time $t$, and
\begin{equation}
M = \begin{pmatrix} -g & 0 & g \\ g & -g & 0 \\
0 & g & -g \end{pmatrix}.  
\label{Mmat}
\end{equation}
The system reaches a steady state for $P_1^*=P_2^*=P_3^*=1/3$.
The transitions $i\rightarrow i+1$
occur with a rough periodicity determined by $g$; that is, the time
evolution of our simple model qualitatively resembles that of the
discretized phase of a generic noisy oscillator with the
intrinsic eigenfrequency set by the value of $g$.

\begin{figure}
\includegraphics[width=7 cm]{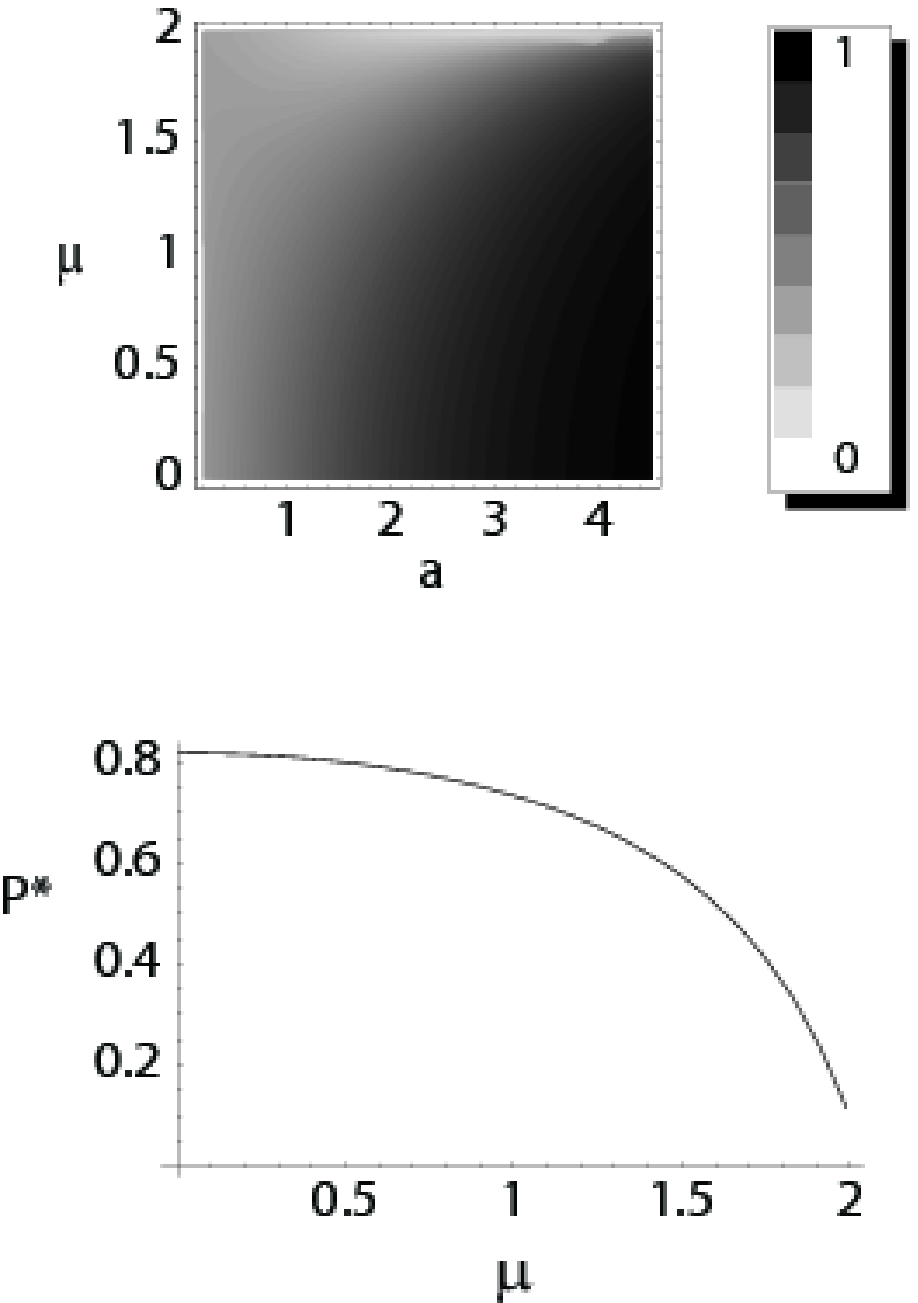}
\caption{The steady state probability $\mathcal{P}_A^*$ that
both units of a dimer are in the same state for a range of $\mu$ and $a$. 
Top: contour image. Bottom: a single curve for $a=2.2$.  In the latter
case, it is clear that as $\mu$ rises, synchronization rapidly decreases.}
\label{dimer1a}
\end{figure}

To study coupled arrays of these oscillators, we couple individual
units by allowing the transition rates of each unit to depend on
the states of the units to which it is connected.  Specifically,
for $N$ identical units we choose the transition rate of a unit $\nu$
from state $i$ to state $j$ as
\begin{equation}
g_{ij} = g \exp\left[{\frac{a(N_{j}-N_{i-1})}{n}}\right]\delta_{j,i+1},
\label{gnu}
\end{equation}
where $\delta$ is the Kronecker delta, $a$ is the coupling parameter,
$g$ is the transition rate parameter, $n$ is the number of oscillators to
which unit $\nu$ is coupled, and
$N_k$ is the number of units among the $n$ that are in state $k$.
Each unit may thus transition to the state ahead or remain in its
current state depending on the states of its nearest neighbors. 
In our earlier work we considered the globally coupled system, $n=N-1$,
and also nearest neighbor coupling in square, cubic, or hypercubic
arrays, $n=2d$ ($d=$ dimensionality).  Here we focus on the globally
coupled array.

Our previous work used a slightly different form of the
coupling~(\ref{gnu}), with $N_{i-1} \to N_i$.  While the
differences in these details do not in any way affect the
characterization of the synchronization transitions, that earlier choice
was numerically advantageous because it led to a phase transition at a lower
critical value $a_c$ of the coupling constant ($a_c=1.5$ in the globally
coupled array) than other choices.  A lower coupling in turn
facilitates numerical integration of
equations of motion because the time step that one needs to use near the
phase transition must be sufficiently small, $dt \ll e^{-a}/g$.  However,
that earlier coupling choice brought with it a result that is
undesirable in our present context (but was of no consequence before).
In our earlier model, as the units become increasingly synchronized
above the transition point, the average
transition rate of a cluster becomes substantially dependent
on the value of $a$; specifically, the transitions and cluster
oscillation frequency slow as $a$ is
increased due to an exponential decrease in the transition probability. 
To cite an explicit example, consider a small subsystem composed of
units which are all in the same state at time $t$ (that is, a cluster
of units which are perfectly synchronized).  The previous form of the
coupling yields an exponentially small transition rate in this case,
and hence the oscillation frequency of this microscopic cluster approaches
zero for high values of $a$.  Since here we specifically wish to analyze
the effects of transition rate disorder, it is desirable to deal with a model
in which the average transition rate of identical synchronized units depends
only on their intrinsic transition rate parameter and not on coupling
strength.  The
form~(\ref{gnu}) reduces simply to the constant $g$ when the coupled
units are perfectly synchronized.  While the critical coupling in this
new version is higher than in our earlier model and hence is numerically
less efficient, no other features of the synchronization transition are
affected.  This in fact supports the desired insensitivity of the
interesting macroscopic features of the model to microscopic modifications.

For a population of $N\to\infty$
identical units in the mean field (globally coupled)
version of this model 
we can replace $N_k/N$ with the probability $P_k$, thereby arriving
at a nonlinear equation for the mean field probability,
$\partial P(t)/\partial t = M[P(t)]P(t)$,  with
\begin{equation}
M[P(t)] = \begin{pmatrix} -g_{12} & 0 & g_{31} \\ g_{12} & -g_{23}
& 0 \\ 0 & g_{23} & -g_{31} \end{pmatrix}.  
\label{Mmatmn}
\end{equation}
\begin{figure}
\includegraphics[width=7 cm]{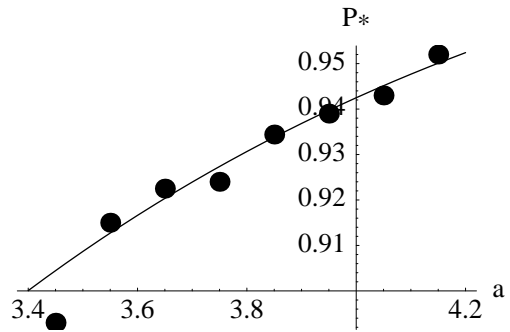}
\caption{The steady state probability $\mathcal{P}_A^*$ that
both units of a dimer are in the same state for
$(\gamma_1,\gamma_2)=(0.5,1.5)$ as $a$ is increased (solid line). 
The points represent simulation results, where $\mathcal{P}_A^*$ is
measured as the fraction of time that both units are fully synchronized.}
\label{dimer2}
\end{figure}
Normalization allows us to eliminate $P_3(t)$ and obtain a closed set
of equations for $P_1(t)$ and $P_2(t)$.  We can then linearize about
the fixed point $(P_1^*,P_2^*)=(1/3,1/3)$, yielding a set of complex
conjugate eigenvalues which determine the stability of this
disordered state.  Specifically, we find that
$2\lambda_\pm/g = (a-3)\pm i\sqrt{3}(1+a)$, eigenvalues that cross
the imaginary axis at $a_c=3$,
indicative of a Hopf bifurcation at this value. Note that the
oscillation frequency of the array at the critical point as given by the
imaginary parts of the eigenvalues is
$\omega=\sqrt{3}g/2$. A
more detailed analysis~\cite{threestate1,threestate2,kuznetsov}
shows the bifurcation to be supercritical.

\begin{figure}
\includegraphics[width=7 cm]{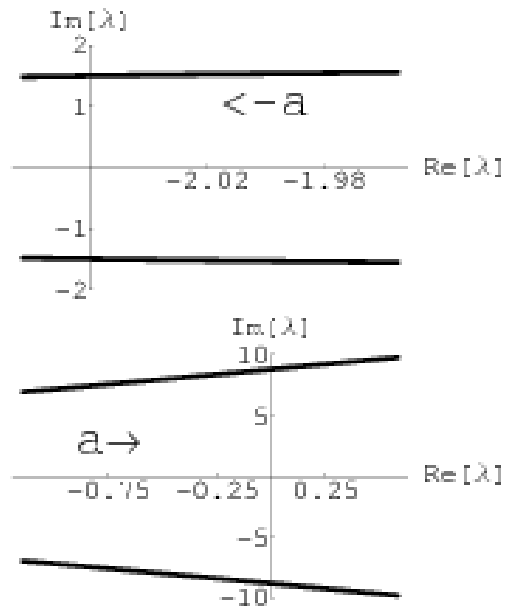}
\caption{The two pairs of complex conjugate eigenvalues for the
dichotomously disordered system, $\mathcal{N}=2$. Top panel:
$(\lambda_-,\lambda_-^*)$. Bottom panel: $(\lambda_+,\lambda_+^*)$.
The coupling constant is in the range $2.9 \le a \le 4.5$, and
the transition rate parameters are chosen to be $(\gamma_1,\gamma_2)=(1,3)$. 
In the bottom panel the critical value of $a$ is $a_c \approx 3.95$.}
\label{plot12N}
\end{figure}

\section{Dimer}
\label{sec3}
Consider first the simplest ``disordered array," namely, a
mutually coupled dimer where one unit is characterized by
$g=\gamma_1$ and the other by $g=\gamma_2$. 
In terms of the states (phases) $S_1$ and $S_2$ of units 1 and 2,
there are 9 possible dimer states, $(S_1,S_2) = (1,1), (1,2), \ldots,
(3,3)$,  but it is not necessary to seek the ensemble distributions
for all of these states in order to decide whether or not the two units
are synchronized.  We can directly write an exact reduced linear
evolution equation for the $3$ states $A$, $B$, and $C$, where $A$
corresponds to any situation where both units are in the same state
[that is, $(S_1,S_2)=(1,1),(2,2)$, and $(3,3)$], state $B$ corresponds
to a situation where unit $1$ is one state ``ahead"
[$(S_1,S_2)=(2,1),(3,2)$, and $(1,3)$], and state $C$ corresponds to
a situation where unit $2$ is one state ``ahead"
[$(S_1,S_2)=(1,2),(2,3)$, and $(3,1)$].  The evolution equation for
these states is the closed linear set 
\begin{equation}
\label{dimer}
\partial \mathcal{P}(t)/\partial t = \mathcal{A} \mathcal{P}(t),    
\end{equation}
with $\mathcal{P}(t)$ the time dependent probability column
vector $(\mathcal{P}_A(t)~\mathcal{P}_B(t)~\mathcal{P}_C(t))^T$ and 
\begin{equation}
\mathcal{A} = \begin{pmatrix} -\gamma_1 - \gamma_2 & b \gamma_2
& b \gamma_1 \\ \gamma_1 & -b^{-1}\gamma_1  - b \gamma_2  
& b^{-1} \gamma_2 \\ \gamma_2 & b^{-1}\gamma_1 & -b
\gamma_1 - b^{-1} \gamma_2 \end{pmatrix},
\end{equation}
and where we have introduced the abbreviation
\begin{equation}
b\equiv e^a.
\end{equation}
This evolution equation is easy to derive from the definition of the
coupling, Eq.~(\ref{gnu}).  For example, when the system is in state
$A$, it can either go to state $B$, which happens when unit $1$ jumps
ahead with transition rate $\gamma_1$, or it can go to state $C$,
which happens when unit $2$ jumps ahead with transition rate $\gamma_2$. 
Similarly, when the system is in state $B$, it can either jump to
state $A$ (when the lagging unit transitions forward) with transition
rate $b\gamma_2 $ or jump to state $C$ (when the leading unit
transitions forward) with transition rate $b^{-1}\gamma_1$.  

\begin{figure}
\includegraphics[width=8 cm]{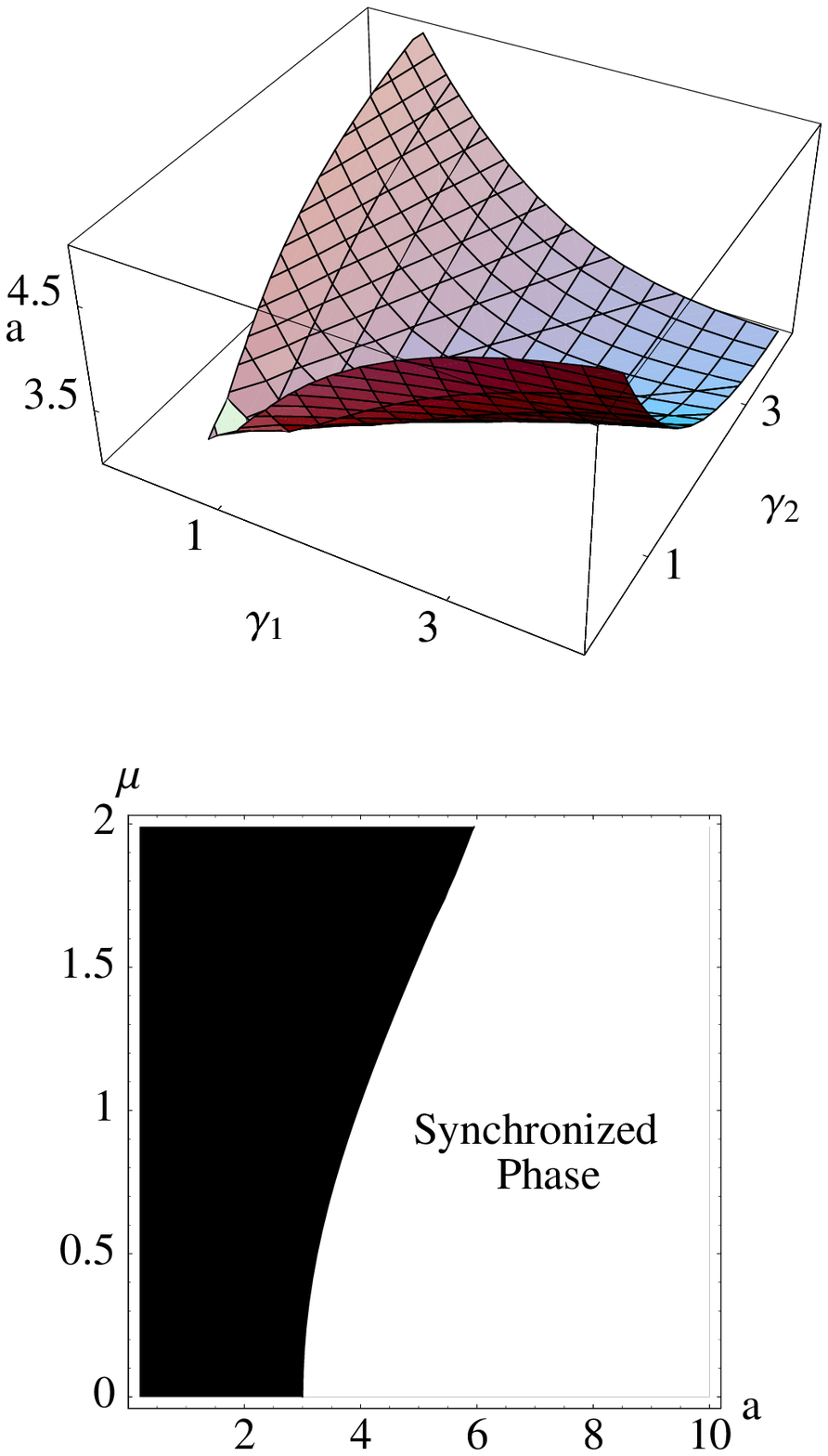}
\caption{(Color online)
Upper panel:
Stability boundary for the dichotomously disordered system.
The contour Re$\lambda_+ = 0$ is plotted in
$(\gamma_1,\gamma_2,a)$ space.  This contour indicates the critical
point, where the Hopf bifurcation occurs and the disordered solution
becomes unstable.  The region above the contour represents the
synchronized phase.  Lower panel: Stability boundary in terms of
relative width parameter.
}  
\label{Contour2Unit}
\end{figure}

With normalization, Eq.~(\ref{dimer}) becomes a $2$-dimensional
equation having steady state solution
\begin{equation}
\begin{aligned}
\label{fpdimer}
\mathcal{P}_A^* & = \frac{b (\gamma_1^2 + b^2 \gamma_1 \gamma_2 +
\gamma_2^2)}{(1+b+b^2) (\gamma_1^2+\gamma_2^2) + (2 + b^3) \gamma_1
\gamma_2 }, \\
\mathcal{P}_B^* & = \frac{b^2 \gamma_1^2 + \gamma_2 (\gamma_1 + \gamma_2)}{(1+b+b^2) (\gamma_1^2+\gamma_2^2) + (2 + b^3) \gamma_1 \gamma_2}. 
\end{aligned}
\end{equation}
The eigenvalues of the two-dimensional matrix obtained from
$\mathcal{A}$ after implementing normalization have negative real
parts for all positive values of the parameters $a$, $\gamma_1$,
and $\gamma_2$, indicating that the fixed points given by
Eq.~(\ref{fpdimer}) are stable.
Hence, the system asymptotically tends to this steady state solution.
We are particularly interested in $\mathcal{P}_A^*$, the probablity
for the system to be synchronized. 
In terms of the single relative width parameter width/mean,
\begin{equation}
\mu\equiv \frac{2(\gamma_1 - \gamma_2)}{(\gamma_1 + \gamma_2)}  
\label{mu}
\end{equation}
($-2\leq \mu \leq 2$), this probability is
\begin{equation}
\mathcal{P}_A^* = \frac{b} {(2+b)}
\left(
   \frac{1+\mu^2\frac{\displaystyle{(2-b^2)}}{\displaystyle{4(2+b^2)}}}{1+\mu^2
   \frac{\displaystyle{b(2+2b-b^2)}}{\displaystyle{4(2+b^2)(2+b)}}}
\right).
\end{equation}
The probability of synchronization for a dimer of identical units
($\mu=0$) is thus $\mathcal{P}_A^* = b/(2+b)=e^a/(2+e^a)$, which
increases with
increasing coupling.  This is the maximal syncrhonization; it is easy to
ascertain that $\mathcal{P}_A^*$ decreases with increasing $\mu^2$, as one
would anticipate.  The full behavior of $\mathcal{P}_A^*$ as a function
of the various parameters is shown in Figs.~\ref{dimer1}-\ref{dimer2}.
The gradual increase in synchronization probability with increasing
coupling turns into a sharp transition as a function of $a$ in the
infinite systems to be considered below.  The decreased synchronization
probability when the frequencies of the two units become more dissimilar
(increasing $\mu$) will also be reflected in the dependence of the
critical coupling on transition rate parameter disorder.

\section{$\mathcal{N}$ Different Transition Rates}
\label{sec4}

Next we consider globally coupled arrays of oscillators that can have
one of $\mathcal{N}$ different transition rate parameters,
$g=\gamma_u$, $u=1,\ldots,\mathcal{N}$. 
To arrive at a closed set of mean field equations for the probabilities
we again go to the limit of an infinite number of oscillators,
$N\to\infty$.  However, we must do so while preserving a finite density
of each of the $\mathcal{N}$ types of oscillators. The probability
vector is now $3\mathcal{N}$-dimensional, 
$P(t)=(P_{1,\gamma_{1}}~P_{2,\gamma_{1}}~P_{3,\gamma_{1}}~\cdots~
P_{1,\gamma_{\mathcal{N}}}~P_{2,\gamma_{\mathcal{N}}}~
P_{3,\gamma_{\mathcal{N}}})^T$. 
The added subscript on the components of $P(t)$
keeps track of the transition rate parameter.
Explicitly, the component $P_{i,\gamma_u}$ is the probability that a unit
with transition rate parameter $g=\gamma_u$ is in state $i$. 
The mean field evolution
for the probability vector is the set of coupled nonlinear differential
equations $\partial P(t)/\partial t = M_{\mathcal{N}}[P(t)]P(t)$, with
\begin{equation}
M_{\mathcal{N}}[P(t)] = \begin{pmatrix} \mathcal{M}_{\gamma_1} & 0 &
\ldots & 0 \\ 0 & \mathcal{M}_{\gamma_2 } & \ldots & 0\\ : &: & : & : \\
0 & \ldots & 0 & \mathcal{M}_{\gamma_{\mathcal{N}}} \end{pmatrix}.  
\label{MmatmnNUnits}
\end{equation}
Here 
\begin{equation}
\mathcal{M}_{\gamma_u} = \begin{pmatrix} -g_{12}(\gamma_{u}) & 0 & g_{31}(\gamma_{u}) \\ g_{12}(\gamma_{u}) & -g_{23}(\gamma_{u})
& 0 \\ 0 & g_{23}(\gamma_{u}) & -g_{31}(\gamma_{u}) \end{pmatrix},  
\label{MBlocks}
\end{equation}
and
\begin{equation}
g_{ij}(\gamma_{u}) = \gamma_{u} \exp\left[a\sum_{k=1}^{\mathcal{N}}
\varphi(\gamma_k) \left(P_{j,\gamma_k}-P_{i-1,\gamma_k}\right)\right]
\delta_{j,i+1}.
\label{gmuNunits}
\end{equation}
The function $\varphi(\gamma_k)$ is the fraction
of units which have a transition rate parameter $g=\gamma_k$. 

As before, probability normalization allows us to reduce this to a
system of $2\mathcal{N}$ coupled ordinary differential equations.  It is
interesting to compare this setup with that of the original Kuramoto
problem with noise, where a continuous frequency distribution is
introduced and the governing equation is a nonlinear partial
differential equation [the Fokker-Planck equation for the
density $\rho(\theta,\omega,t)$]~\cite{acebron}.  The discretization
of phase in our model results instead in a set of $2\mathcal{N}$
coupled nonlinear ordinary differential equations.

\begin{figure}
\includegraphics[width=8 cm]{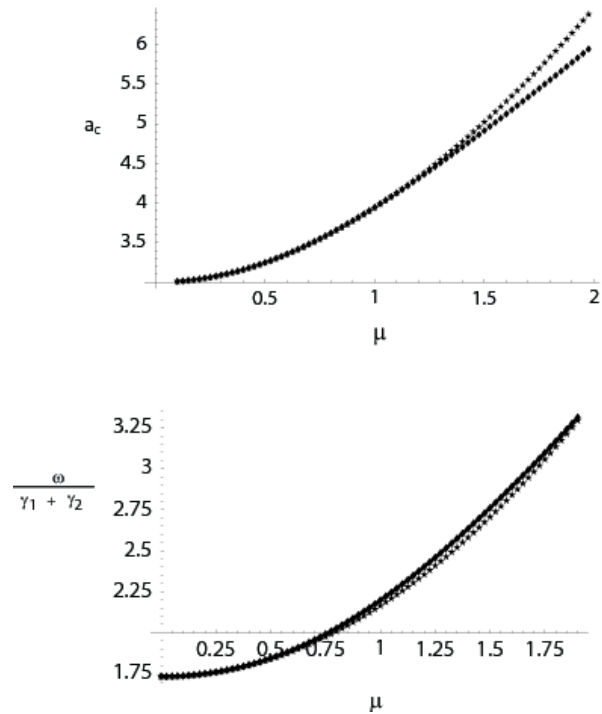}
\caption{Upper panel: Critical coupling $a_c$ as a function of $\mu$
for a dichotomous ($\mathcal{N}=2$) array of globally coupled
oscillators. The two curves represent the exact relationship
(lower curve) and the
small $\mu$ approximation (upper curve), respectively. 
Lower panel: The frequency of synchronous oscillation at the transition
point.  Lower curve is the approximation as predicted by ${\rm
Im}\lambda_+$, upper curve is the exact result.}
\label{plot22N}
\end{figure}

While it is nevertheless still difficult to solve these 
equations even for small $\mathcal{N}$, we can
linearize about the disordered state $P(t) = (1/3~1/3~\ldots ~1/3)^T$
and arrive at a $2\mathcal{N}\times 2\mathcal{N}$ Jacobian of
the block matrix form
\begin{equation}
J=\begin{pmatrix}  \mathcal{J}_1(\gamma_1) & \mathcal{J}_2(\gamma_1)
& \mathcal{J}_2(\gamma_1) & \ldots & \mathcal{J}_2(\gamma_1) \\ 
\mathcal{J}_2(\gamma_2) & \mathcal{J}_1(\gamma_2) & \mathcal{J}_2(\gamma_2)
& \ldots & \mathcal{J}_2(\gamma_2) \\ : & : & : & : & : \\
: & : & : & : & : \\ \mathcal{J}_2(\gamma_N) & \mathcal{J}_2(\gamma_N)
& \ldots & \mathcal{J}_2(\gamma_N) & \mathcal{J}_1(\gamma_N)  \end{pmatrix}.
\label{jacobN}
\end{equation}
The blocks $\mathcal{J}_1(g)$ and $\mathcal{J}_2(g)$ are given by:
\begin{equation}
\mathcal{J}_1(g)=\begin{pmatrix} -2 g & -g - ag/N \\
g+  ag/N) & -g + a g/N \end{pmatrix} 
\end{equation}
and
\begin{equation}
\mathcal{J}_2(g)=\begin{pmatrix} 0 & -ag/N\\
a g/N & a g/N \end{pmatrix} .
\end{equation}
While we explore this in more detail below only for small $\mathcal{N}$,
we note that in general the Jacobian~(\ref{jacobN}) has $\mathcal{N}$ pairs of
complex conjugate eigenvalues, only \emph{one pair of which seems to
have a real part that becomes positive} with increasing coupling constant
$a$. This implies that there is a \emph{single} transition to synchrony
even in the presence of the transition rate disorder that we have introduced
here.  We go on to confirm this behavior for $\mathcal{N}=2$,
$3$, and $4$.

\subsection{Two transition rate parameters}
\label{sub41}

For the $\mathcal{N}=2$ case, the four eigenvalues
$(\lambda_+,\lambda_+^*,\lambda_-,\lambda_-^*)$ 
of the Jacobian can
be determined analytically.  We find
\begin{equation}
\begin{aligned}
\label{eigenanalytical}
\frac{{\rm Re} \lambda_{\pm}}{\gamma_1+\gamma_2} &= \frac{1}{8} \left[a-6 \pm
B(a,\mu)\cos\left(C(a,\mu)\right)\right],\\
\frac{{\rm Im} \lambda_{\pm}}{\gamma_1+\gamma_2} &= \frac{1}{8}
\left[\sqrt{3}(a+2) \pm B(a,\mu)\sin \left(
C(a,\mu)\right)\right],
\end{aligned}
\end{equation}
where
\begin{equation}
\begin{aligned}
B(a,\mu)&\equiv \sqrt{2} \left[a^4 - 6 a^2 \mu^2 + 3 \mu^4
(a^2+3)\right]^{1/4},\\
C(a,\mu)&\equiv \frac{1}{2} \tan^{-1} \left(\frac{- \sqrt{3} (a^2
-(a+3) \mu^2)}{a^2+3(a-1) \mu^2}\right).
\end{aligned}
\end{equation}
Aside from an overall factor $(\gamma_1+\gamma_2)$, 
Eqs.~(\ref{eigenanalytical}) depend only on the relative width variable
as defined in Eq.~(\ref{mu}),
and therefore the critical coupling $a_c$ depends only on $\mu$. 
As illustrated in Fig.~\ref{plot12N}, one pair of 
eigenvalues crosses the imaginary axis at a critical value $a=a_c$,
but the other pair shows no qualitative change as
$a$ is varied.  While this figure shows only the particular 
transition rate parameter values $(\gamma_1,\gamma_2)=(1,3)$, the
qualitative features of these eigenvalues remain similar for the
entire range of positive parameters. The upper panel of
Fig.~\ref{Contour2Unit} depicts the contour Re$\lambda_+ = 0$ in
$(\gamma_1,\gamma_2,a)$ space; this contour represents the critical
surface and thus separates the synchronous and disordered phases.   

The critical coupling is the value of $a$ at which Re$\lambda_+=0$
(Re$\lambda_-$ does not vanish for any $a$). It is easy to ascertain
that Im$\lambda_+$ does not vanish at $a_c$, so that the critical point
is a Hopf bifurcation.  Furthermore, it is clear from
Eq.~(\ref{eigenanalytical}) that $a_c$ depends only on the relative
width parameter $\mu$, and it is also straightforward to
establish that $a_c$ increases with increasing $\mu$, that is, a
stronger coupling is necessary to overcome increasingly different values
of $\gamma_1$ and $\gamma_2$ (see lower panel of
Fig.~\ref{Contour2Unit}).  Note, however, that the dependence 
on $\mu$ implies that it is not just the difference in transition rates
but the \emph{relative} difference or percent difference relative to the
mean transition rate that is the determining factor in how strong
the coupling
must be for synchronization to occur.  A small-$\mu$ expansion leads to
an estimate of $a_c$ to $O(\mu^2)$,
\begin{equation}
\label{acapprox}
a_c \approx \frac{1}{8} \left(12 + 3 \mu^2 + \sqrt{3}
\sqrt{(12+\mu^2)(4+3 \mu^2)}\right),
\end{equation} 
a result that exhibits these trends explicitly.  The upper panel in
Fig.~\ref{plot22N}
shows that this estimate is remarkably helpful even when $\mu^2$ is not
so small.

The frequency of oscillation of the synchronized system at the transition
is given by $\omega=\lim_{a\to a_c}$Im$\lambda_+$.  From
Eq.~(\ref{eigenanalytical}) it follows that $\omega$ depends on 
$(\gamma_1+\gamma_2)$ as well as $\mu$.  The small-$\mu$ expansion leads
to the estimate
\begin{equation}
\label{omegaapprox}
\omega 
= Im (\lambda_{\pm}) |_{a \to a_c} \approx \frac{1}{4}
\sqrt{3} (\gamma_1+\gamma_2) (4 + \mu^2),
\end{equation}
which works exceedingly well for all $\mu$ (see Fig.~\ref{plot22N}).

\begin{figure}
\includegraphics[width=8 cm]{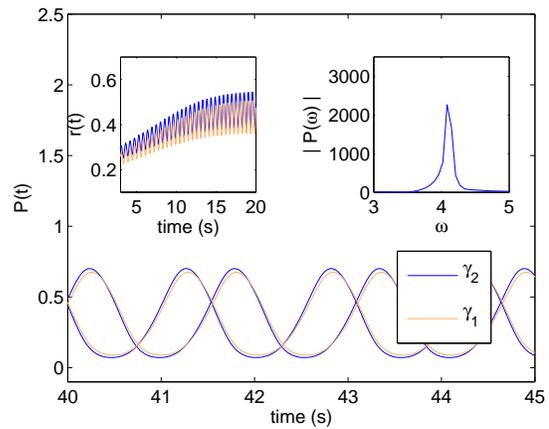}
\caption{(Color online)
The components $P_{1,\gamma_1}$ and $P_{2,\gamma_1}$(two lighter or
brown curves),
and $P_{1,\gamma_2}$ and $P_{2,\gamma_2}$(two darker or blue curves), of
the vector $P(t)$ vs time for $\gamma=1,~ \Delta=0.125$, with
$a = 3.15$, which is above the critical value $a_c \approx 3.02$
predicted by linearization.  The left inset shows the order parameter
$r(t)$ as it approaches its long-time limit. The right inset
shows the frequency spectrum of a component of $P(t)$.
The spectrum has a dominant peak near $\omega\approx 4$, and is expected
to approach the frequency $\omega \approx 3.5$ predicted by linearization
as we approach the transition point $a \to a_c$.} 
\label{plot72N}
\end{figure}

\begin{figure}
\includegraphics[width=8 cm]{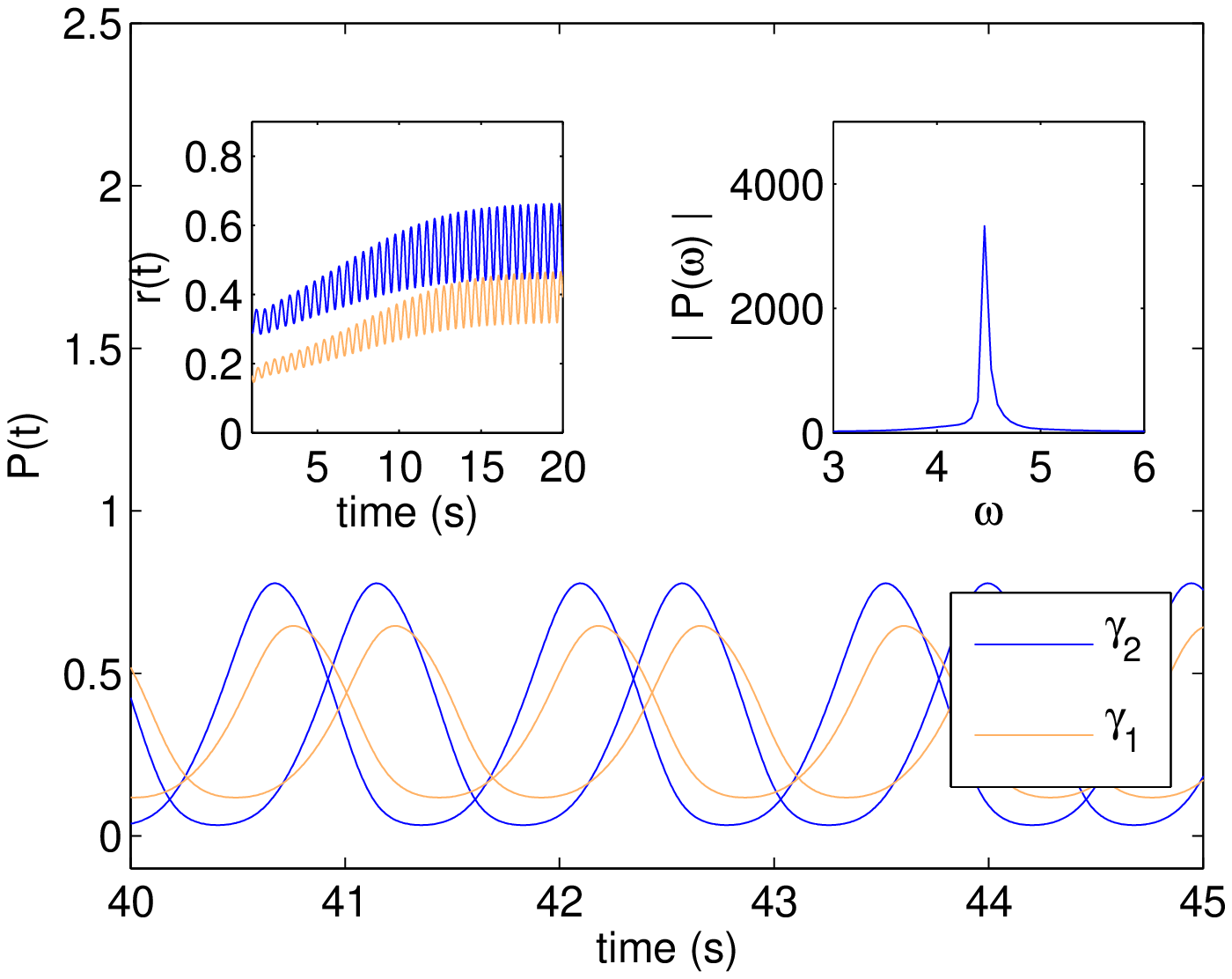}
\caption{(Color online)
The components $P_{1,\gamma_1}$ and $P_{2,\gamma_1}$(two lighter or
brown curves),
and $P_{1,\gamma_2}$ and $P_{2,\gamma_2}$(two darker or blue curves), of
the vector $P(t)$ vs time for $\gamma=1,~ \Delta=0.625$, with
$a = 3.5$, which is above the critical value $a_c \approx 3.39$
predicted by linearization.  The left inset shows the order parameter
$r(t)$ as it approaches its long-time limit. The right inset
shows the frequency spectrum of a component of $P(t)$.
The spectrum has a dominant peak near $\omega\approx 4.4$, and is expected
to approach the frequency $\omega \approx 3.8$ predicted by linearization
as we approach the transition point $a \to a_c$.} 
\label{plot82N}
\end{figure}

\begin{figure}
\includegraphics[width=8 cm]{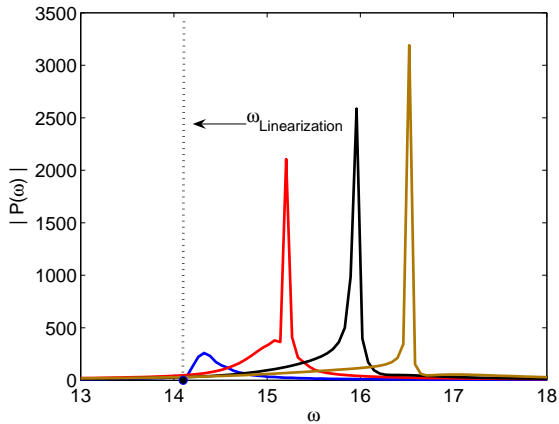}
\caption{(Color online)
The frequency $\omega$ of macroscopic oscillations approaches
the value predicted by linearization as $a\to a_c$. Here we have used
$\gamma_1=3.5$ and $\gamma_2=4.5$ so that $\gamma=4$ and $\Delta=1$.
The critical coupling constant is $a_c=3.06$. From darkest or blue to
lightest or brown: $a=3.07$, $3.10$, $3.15$, and $3.20$. }
\label{plotfThresh}
\end{figure}

\begin{figure}
\includegraphics[width=8 cm]{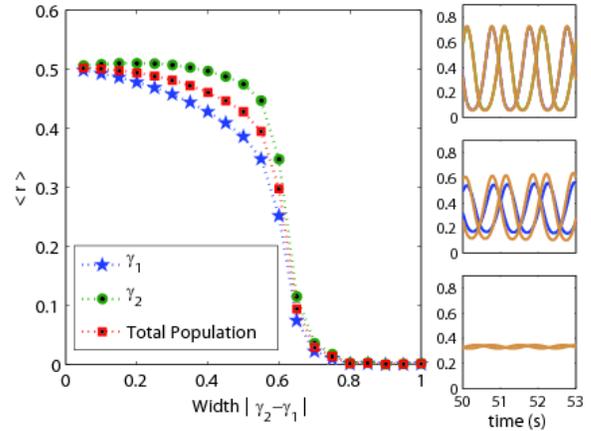}
\caption{(Color online)
Time-averaged order parameter $r$ in the long-time limit vs
$\Delta$ for the individual oscillator populations characterized
respectively by the transition rate parameter $\gamma_1$ (stars) and
$\gamma_2$ (circles), and
for the entire mixed array (squares).  The insets show the
time evolution of the probability vector components $P_{1,\gamma_1}$
and $P_{2,\gamma_1}$ (lighter or brown curves) and $P_{1,\gamma_2}$ and
$P_{2,\gamma_2}$ (darker or blue curves) for widths 0.05 (upper inset), 0.5
(middle inset), and 0.9 (lower inset).  Some of the curves are not
visible because they are so perfectly superimposed.
While the degree of synchronization varies within each population,
the critical width for de-syncrhonization is the same for both, as
predicted by linearization.
The coupling constant for all cases is $a=3.2$ and the
average transition rate parameter $\gamma=1.5$.
}
\label{plot92N}
\end{figure}

To check the predictions of our linearization procedure, we numerically
solve the nonlinear $\mathcal{N}=2$ mean field equations.  In agreement
with the structure of the linearized eigenvalues, all components
of $P(t)$ synchronize to a common frequency as the phase boundary in
$(\mu, a)$-space) is crossed.  Interestingly, the numerical solutions also
give us insight into the amplitude of the oscillations; that is,
they allow us to explore the relative ``magnitude" of synchronization
within the two populations. 
As we will see, the two populations indeed oscillate with the same
frequency, but with amplitudes and ``degrees of synchronization" that
can be markedly different.  
Consider the order parameter $r(t)$ given by
\begin{equation}
r(t) \equiv \frac{1}{N} \left \lvert \sum_{\nu=1}^N e^{i \phi_\nu}
\right \rvert.
\label{orderp}
\end{equation}
Here $\phi_\nu$ is the discrete phase 2$\pi$($k$-1)/3
for state $k \in \lbrace 1,2,3 \rbrace$ at site $\nu$.  For phase
transition studies, one would likely average this quantity over time 
in the long time limit, and also over independent trials.  For our
purposes here,
though, we find the time-dependent form more convenient.  In the mean
field case, where we solve for probabilities to be in each state,
the order parameter is easily calculated by writing the average in
Eq.~(\ref{orderp}) in terms of these probabilities rather than as a
sum over sites.

\begin{figure}
\includegraphics[width=8 cm]{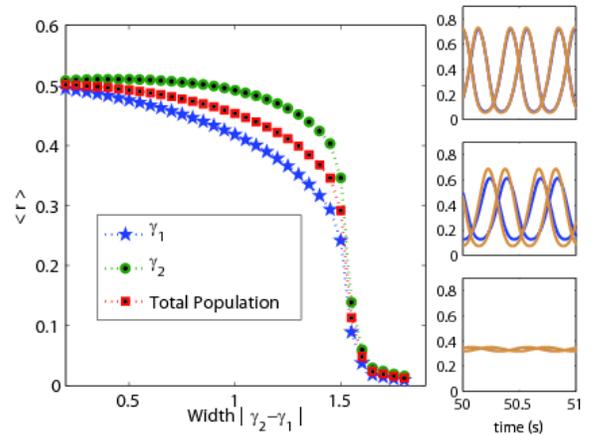}
\caption{(Color online)
Same as Fig.~\ref{plot92N} but with $\gamma=3.5$ and widths 0.1
(upper inset), 1.3 (middle), and 1.8 (lower).}
\label{plot102N}
\end{figure}

In subsequent figure captions we introduce the notation $\gamma\equiv
(\gamma_1+\gamma_2)/2$ (average transition rate parameter),
and the difference $\Delta\equiv
|\gamma_2-\gamma_1|$ (note that $\mu=\Delta/\gamma$). 
As shown in Figs.~\ref{plot72N}-\ref{plot102N}, the predictions
of linearization accurately describe the onset of macroscopic
synchronization and provide an estimation of the frequency of
these oscillations near threshold (see Fig.~\ref{plotfThresh}). 
Specifically, Figs.~\ref{plot72N} and~\ref{plot82N} show the
macroscopic oscillations for coupling $a$ above threshold.
All the oscillators, regardless of their intrinsic transition rate parameter,
oscillate exactly in phase, but
the degree of synchronization is greater in the population
with the larger $\gamma_i$ (here $\gamma_2$), as evidenced by
the unequal amplitude of the components of $P(t)$ for the two
populations. The
``greater degree of synchronization" is also apparent in the order
parameter $r(t)$ shown in the insets, which is larger for the
oscillators with the higher intrinsic transition rate.   These results
support the notion that populations with higher transition rate parameters
in some sense synchronize more readily.  The figures also show the
frequency spectrum of any component of $P(t)$.  The peak occurs at the
frequency of oscillation of the synchronized array.  As $a\to a_c$ this
frequency approaches the value Im$\lambda_+$ predicted by linearization,
as shown in Fig.~\ref{plotfThresh}.

Figures~\ref{plot92N} and~\ref{plot102N} illustrate the sudden
de-synchronization (at fixed $a$ and $\gamma$) accompanying an increase
in the difference $\Delta$.  This behavior is reminscent of
that of the original Kuramoto oscillators, which become
disordered as the width of the frequency distribution characterizing
the population exceeds some critical value.  The insets show 
the components  of $P(t)$ and confirm that both populations undergo the
de-syncrhonization transition at the same critical value of the
difference $\Delta$.  Comparing the two figures, we see that the system
with a higher average transition rate parameter
(Fig.~\ref{plot102N}) can withstand a
larger difference $\Delta$ before de-syncrhonization, again confirming
our earlier observations.

\begin{figure}
\includegraphics[width=8 cm]{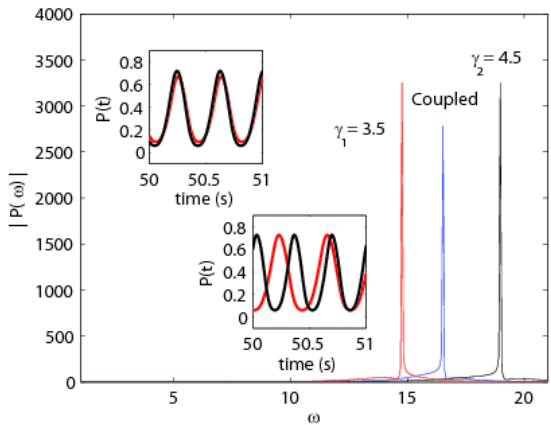}
\caption{(Color online) Frequency spectra of the numerical solutions
to the mean
field equations ($\mathcal{N}=2$) show that above critical coupling,
$a=3.2>a_c=3.06$,
synchronization occurs at a frequency closer to the lower of the
two population frequencies.  Top left inset: $P_{1,\gamma_1=3.5}$
and $P_{1,\gamma=4.5}$ when all units are globally coupled. 
Bottom right inset: the same curves for populations that
are uncoupled from one another (but still globally coupled
\emph{within} each population).}
\label{2popsf}
\end{figure}

One last point to consider is the relation between the frequency of
oscillation of the synchronized array above $a_c$ and the frequencies of
oscillation of the two populations if they were decoupled from one
another.  As coupling
increases, the oscillation frequency $\omega$ moves closer to that of
the population with the lower transition rate parameter.  This is
illustrated in Fig.~\ref{2popsf} for the
same parameters used in Fig.~\ref{plotfThresh}.

\begin{figure}
\includegraphics[width=8 cm]{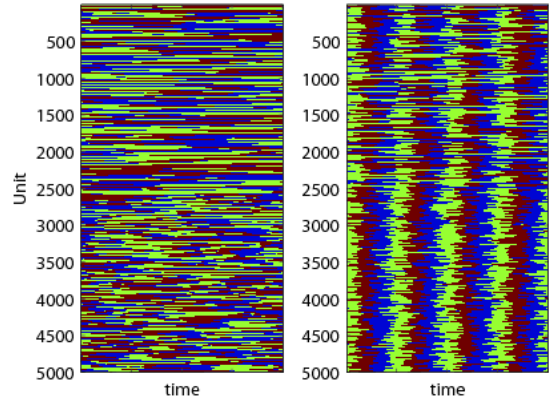}
\caption{(Color online) Long-time snapshots of a globally coupled
system above and
below threshold. 
In both cases, $(\gamma_1,\gamma_2)=(0.5,1.5)$.  On the left, $a=3.5<a_c$
while on the right, $a=4.1>a_c$. 
In both cases, all units are globally coupled. 
For visualization purposes, the plot is arranged so that population
$\gamma_1$ consists of the first $2500$ units (the top), while
population $\gamma_2$ consists of the second $2500$ units (the bottom). 
Global synchrony emerges for $a>a_c$.  In addition, the population
with the higher transition rate parameter is more synchronized.}
\label{plot2NSim}
\end{figure}

Finally, a visually helpful illustration of these behaviors is obtained
via a direct simulation of an array with a dichotomous population of
oscillators.  Since our oscillators are globally (all-to-all) coupled,
the notion of a spatial distribution is moot, and for visulatization
purposes we are free to arrange the populations in any way we wish.  In
Fig.~\ref{plot2NSim} we display an equal number of $\gamma_1$ and
$\gamma_2$ oscillators and arrange the total polulation of $N=5000$ so
that the first 2500 have transition rate parameter $\gamma_1$ and the
remaining 2500 have transition rate parameter $\gamma_2$.  In this
simulation we have chosen $\gamma_1=0.5$ and $\gamma_2=1.5$, so that
$\gamma=\Delta=1$.  Although $N=5000$ is not infinite, it is large
enough for this array to behave as predicted by our mean field theory.
The left panel shows snapshots of the phases (each phase is indicated by
a different color) when $a<a_c$ and the phases are random.  The right
panel shows the synchronized array when $a>a_c$.  Clearly, all units are
synchronized in the right panel, but the population with the higher
transition rate parameter (lower half) shows a higher degree of
synchronization (higher $P(t)$) as indicated by the
intensity of the colors or the gray scale.

\begin{figure}
\includegraphics[width=7 cm]{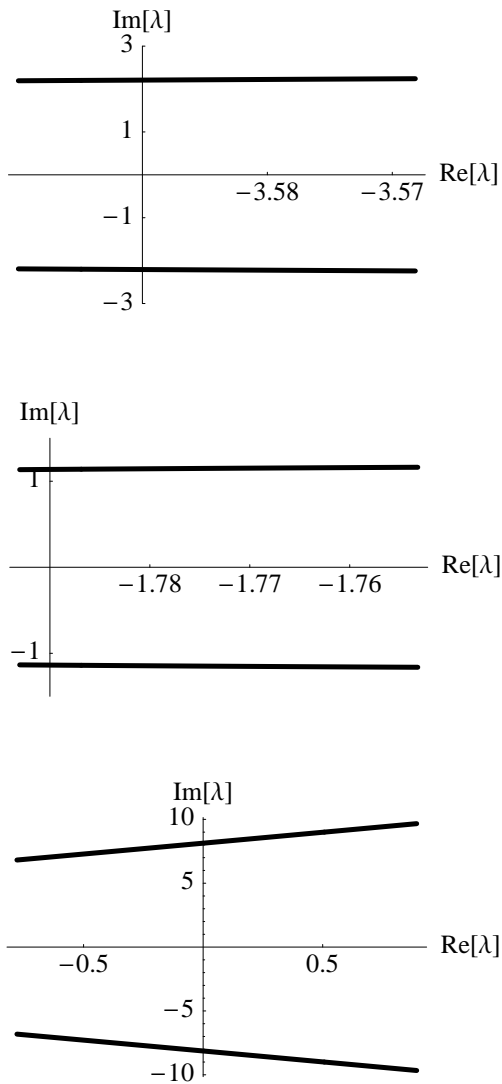}
\caption{$\mathcal{N}=3$ case:  The three pairs of complex
conjugate eigenvalues are plotted in the complex plane for
$2.9 \le a \le 4.5$.  The transition rate parameters are chosen to
be  $(\gamma_1,\gamma_2,\gamma_3)=(1,2,3)$.  In the bottom panel
the critical value of $a$ is $a_c \approx 3.6$.}
\label{plot13N}
\end{figure}

\subsection{$\mathcal{N} =3$ and $\mathcal{N}=4$}
\label{sub42}

\begin{figure}
\includegraphics[width=7 cm]{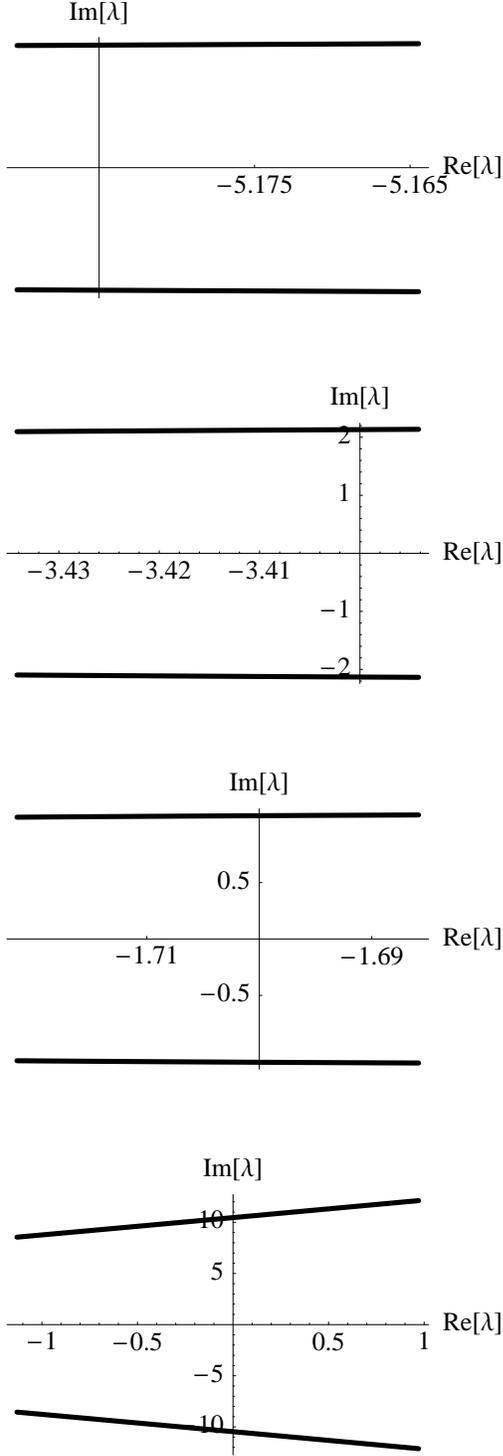}
\caption{$\mathcal{N}=4$ case:  The four pairs of complex conjugate
eigenvalues are plotted in the complex plane for
$2.9 \le a \le 4.5$.  The transition rate parameters are chosen
to be  $(\gamma_1,\gamma_2,\gamma_3,\gamma_4)=(1,2,3,4)$.  In the
bottom panel, the critical value of $a$ is $a_c \approx 3.75$.}
\label{plot14N}
\end{figure}

We can carry out this analysis, albeit not analytically (at least in
practice), for any $\mathcal{N}$.  We have explored the cases
$\mathcal{N}=3$ and $4$. In both cases there appears to be 
only one pair of eigenvalues whose real parts can become positive,
suggesting that synchronization occurs all at once and not in one
population at a time (Figs.~\ref{plot13N} and \ref{plot14N}). 
This occurs no matter the distribution of the 3 or 4 transition rate
parameters.  For example, in the $\mathcal{N}=4$ case we have compared in
some detail the cases where the four transition rates are equidistant and 
where they are pairwise much closer than the separation between the
highest and lowest.  In both cases there is a single transition to
synchrony, albeit not at exactly the same value of $a_c$, indicating a
more complex dependence on the transition rate parameter distribution than
just via its mean and width.  Furthermore, the basic trends of the
dichotomous case broadly carry over, mainly in that the critical value
$a_c$ increases when the width of the distribution increases relative
to the mean (as one would expect).  On the other hand, the inclusion of
more transition rates within a given range leads to a lowering of the
critical coupling.  Thus, for example, the mean transition rate $\gamma$ and
the width $\Delta$ are the same in the cases shown in Figs.~\ref{plot12N}
and \ref{plot13N} ($\gamma=\Delta=2$), and yet $a_c$ is higher
in the former
(3.95 for $\mathcal{N}=2$) than in the latter (3.6 for $\mathcal{N}=3$). 
Still, the mean and width of the distribution provide a
rough qualitative assesment of the behavior, particularly for the case
of a uniform distribution, which we study below.  

\section{Uniform Distribution of Transition Rate Parameters}
\label{sec5}

We now turn to globally coupled arrays where the transition rate
parameter $g$ for each unit is chosen from a uniform distribution over a
finite interval, $\varphi(g)$.  While it is difficult to make
direct analytical progress in this general case, the earlier dimer
analysis and the arrays of $\mathcal{N}=2,3,4$ different populations
of units provide a framework for understanding the properties of these
more general systems.  In particular, the earlier results
suggest that this ``more disordered" system may also display a single
transition to synchronization. 
To explore these and other features in more detail,
we simulate $N=5000$ globally connected
units characterized by the transition rate
parameter distribution $\varphi(g)$,
and we make several observations.  Firstly, we do observe a single
transition to macroscopic synchronization.  Secondly,
as suggested by the dichotomous case, synchronization appears more
readily (that is, for a lower value of $a$) if the distribution
$\phi(g)$ has a larger mean and smaller width.  When the mean and width
are varied independently, the qualitative trends from the dichotomous
case are observed here as well.
Thirdly, while synchronization in this system is again governed
primarily by the mean and width of the distribution $\varphi(g)$, 
the critical value $a_c$ is considerably lower than that of the finite
$\mathcal{N}$ systems with the same mean and width (as expected).

\begin{figure}
\includegraphics[width=7 cm]{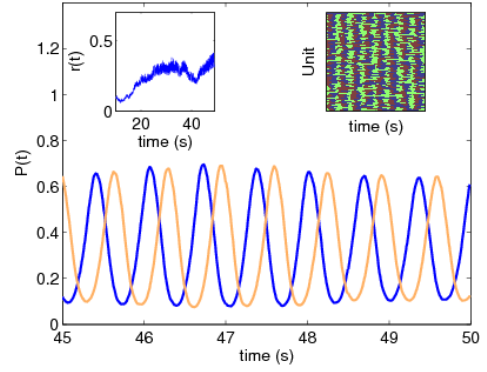}
\caption{(Color online)
The probability that the synchronized array is in state 1
(lighter or brown) and state 2 (darker or blue) as a function of time
for a uniform distribution $\varphi(g)$ on the interval
$[1.5,2.5]$ and coupling parameter $a=3$.
Insets show the order parameter $r(t)$ as well as time resolved
snapshots of the system. 
}
\label{Sim1}
\end{figure}

\begin{figure}
\includegraphics[width=7 cm]{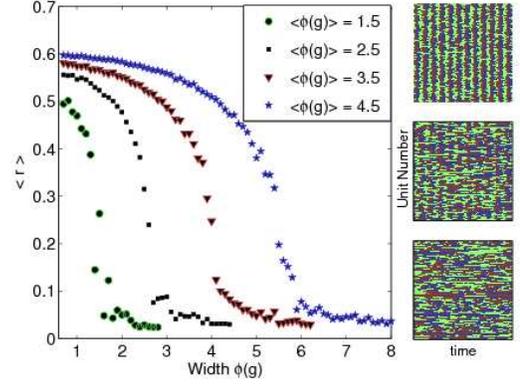}
\caption{(Color online)
As the width of the $\phi(g)$ distribution increases,
a critical width is reached beyond which synchronization is destroyed. 
The coupling is chosen to be $a=3.2$, and the four curves represent
the steady state, time-averaged order parameter for distributions
with different means.  As the mean of the $\phi(g)$ distribution
increases, the transition to disorder occurs at a greater width. 
The insets at the right show the long-time behavior of an
entire population of mean transition rate
parameter 3.5 (corresponding to the
triangle order parameter data) and widths of $0.6$, $4.0$, and $6.2$.}
\label{Sim2}
\end{figure}

\begin{figure}
\includegraphics[width=7 cm]{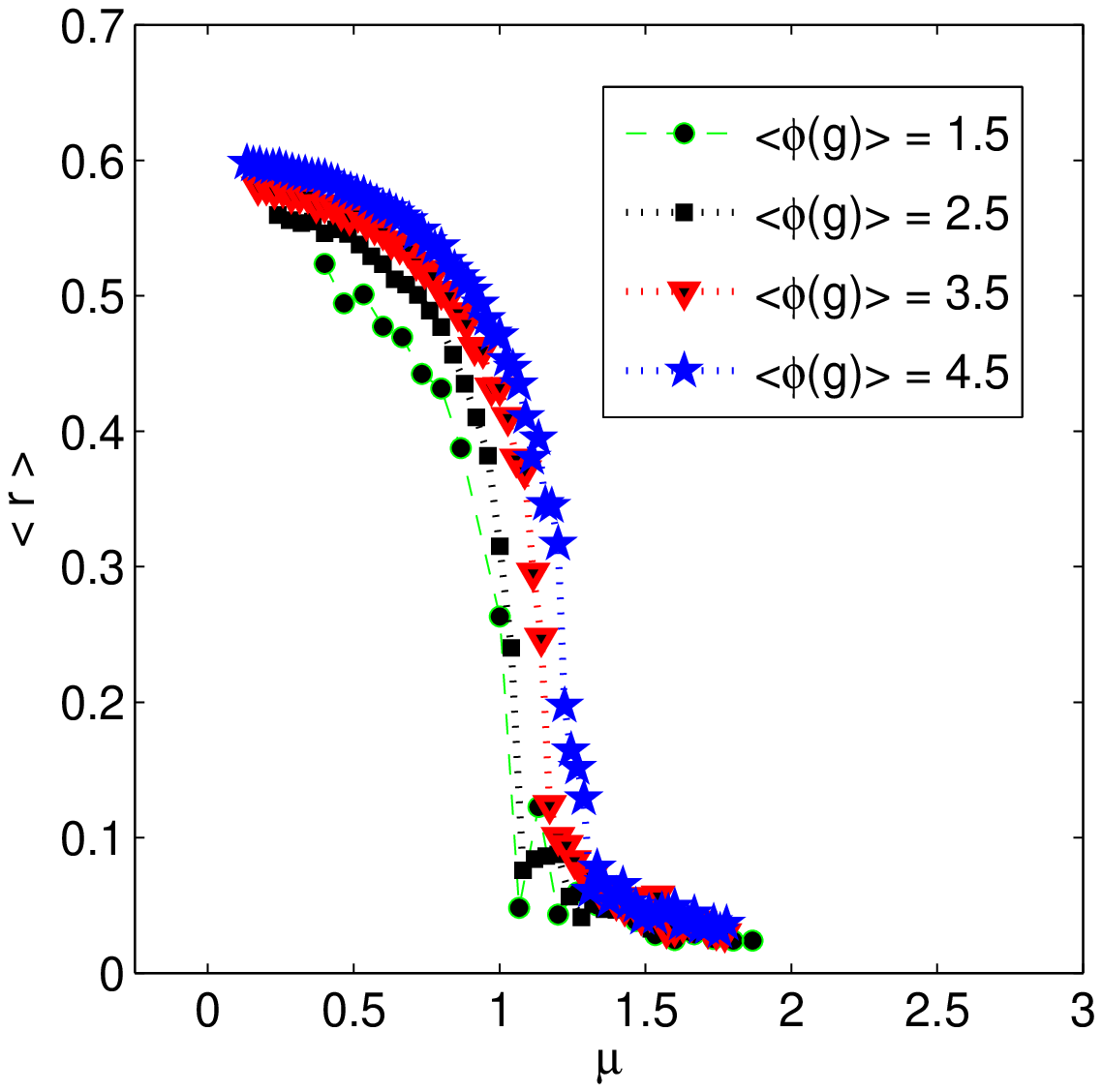}
\caption{The data of Fig.~\ref{Sim2} against the relative
width $\mu$.}
\label{Sim3}
\end{figure}
      
Two examples of our simulation results are shown in
Figs.~\ref{Sim1} and~\ref{Sim2}.  In Fig.~\ref{Sim1} we present the
first two components of the 3-dimensional vector $P(t)$ whose components
$P_i(t)$ represent the probability that all units of the entire
synchronized array are in state $i$. The probabilities $P_1(t)$ and
$P_2(t)$ oscillate in time with essentially constant amplitude and a
constant relative phase, indicating global synchronization.  The upper
left inset shows the order parameter $r(t)$ and the upper right inset the
time resolved snapshot of the system, both indicating a high degree
of synchronization.  Note that the coupling parameter $a=3$ in the figure is 
below the critical value $a_c=3.2$ for the dichotomous case with the
same mean and width.

Figure ~\ref{Sim2} shows the steady state time-averaged order
parameter $r$ at constant $a$ as the width of the $\phi(g)$
distribution is increased for a fixed mean.  Similar to the
$\mathcal{N}=2$ population case, synchronzation is destroyed as
the width eclipses some critical value, and that value increases as
the mean of the distrubtion increases.  In Fig.~\ref{Sim3} we plot the
data from Fig~\ref{Sim2} as a function of the relative width
parameter $\mu$.  Recalling that for the dichotomous array as well as
for the dimer synchronzation at a given $a$ depends only on $\mu$,
we might expect that the transition point $\mu_c$ (at constant $a$)
is not significantly mean-dependent, even when there is a distribution
of transition rate parameters.  In fact, we can see that the curves
approximately collapse onto one curve, indicating that the relative
width $\mu$ provides a useful control parameter for predicting
synchronization.  Hence, the predictions of the linearization analysis
for the $\mathcal{N}=2$ case provides qualitatively insight into the
behavior of the disordered population.

\section{Discussion}  
\label{sec6}

We have presented a discrete model for globally coupled stochastic
nonlinear oscillators with a distribution of transition rate parameters. 
Our model exhibits a range of interesting dynamical behavior, much of which
mimics the qualitative features of the canonical Kuramoto
oscillator~\cite{kuramoto}, but with a mathematically and numerically
considerably more tractable model.  Since our phase variable is
discrete (whereas the phase variable in the canonical problem is
continuous), a distribution of $\mathcal{N}$ different transition rates in
our array leads to a set of $2\mathcal{N}$ coupled nonlinear
\emph{ordinary} differential equations instead of a single
\emph{partial} differential equation for the probability
distributions of interest.  Linearization of our model around the
critical point leads to a problem which at least for small $\mathcal{N}$
(specifically, for the dichotomous disorder case) becomes
analytically tractable.  Distributions involving
a large finite number of transition rate parameters, while not
easily amenable to
analytic manipulation even upon linearization, reduce to a simple matrix
algebra problem.  For any distribution of transition rate parameters, even 
continuous, the model is in any case readily amenable to  numerical
simulation.

Our most salient conclusion is that such disordered globally coupled
arrays of oscillators, even in the face of transition rate parameter disorder,
undergo a single transition to macroscopic synchronization. Furthermore,
we have shown that the critical coupling $a_c$ for synchronization 
depends strongly (but not exclusively) on the width $\Delta$ and
mean $\gamma$ of the transition rate parameter distribution, specifically via
the relative width $\mu=\Delta/\gamma$.  This general feature is already
apparent in the synchronization behavior of
a dimer of two oscillators with transition rate parameters $\gamma_1$ and
$\gamma_2$.  An infinite array of two populations of oscillators, 
one with transition rate parameter $\gamma_1$ and the other with $\gamma_2$,
displays a Hopf bifurcation, with $a_c$ determined 
solely by $\mu$.  While a quantitative prediction of synchronization on
the basis of the relative width is not possible in all cases, it
does determine
qualitative aspects of the transition for more complex transition rate
parameter distributions.  We have explored this assertion for
arrays with $\mathcal{N}=2,~3,$ and $4$ and with a uniform distribution
of transition rates over a finite interval, and expect it be appropriate for
other smooth distributions as well. 

A number of further avenues of investigation based on our stochastic
three-state phase-coupled oscillator model are possible.  For example,
we could explore the effects of transition rate disorder in locally
coupled arrays
whose behavior we have fully characterized for identical
oscillators~\cite{threestate1,threestate2}.  It would be interesting to
explore the consequences of disorder in the coupling parameter $a$.
Finally, we note that a two-state version of this model (which of course
does not lead to phase synchronization as discussed here) has recently
been shown to accurately capture
the unique statistics of blinking quantum dots~\cite{grigolini}.
Such wider applicability of the model, together with its 
analytic and numerical tractability, clearly opens the door to a number
of new directions of investigation.

\section*{Acknowledgments}
This work was partially supported by the National Science Foundation
under Grant No. PHY-0354937.


\begin{thebibliography}{99}

\bibitem{strogatz} S. H. Strogatz, {\em Nonlinear Dynamics and Chaos}
(Westview Press, 1994).

\bibitem{winfree} A. T. Winfree, J. Theor. Biol. {\bf 16}, 15 (1967).

\bibitem{kuramoto} Y. Kuramoto, {\em Chemical Oscillations, Waves,
and Turbulence} (Springer, Berlin, 1984).

\bibitem{strogatz2} S. H. Strogatz, Physica D {\bf 143}, 1 (2000).

\bibitem{pikovsky} A. Pikovsky, M. Rosenblum, J. Kurths,
{\em Synchronization: A Universal Concept in Nonlinear Science}
(Cambridge University Press, Cambridge, 2001).

\bibitem{lutz} T. Prager, B. Naundorf, and L. Schimansky-Geier, Physica
A {\bf 325}, 176 (2003).

\bibitem{local}  H. Sakaguchi, S. Shinomoto, and Y. Kuramoto, Prog.
Theor.  Phys. {\bf 77}, 1005 (1987); H. Daido, Phys. Rev. Lett.
{\bf 61}, 231 (1988); S. H. Strogatz and R. E. Mirollo, J. Phys. A
{\bf 21}, L699 (1988); {\it idem}, Physica D {\bf 31}, 143 (1988);
H. Hong, H. Park, and M. Choi,  Phys.  Rev. E {\bf 71}, 054204 (2004).

\bibitem{threestate1} K. Wood, C. Van den Broeck, R. Kawai, and K. Lindenberg, 
Phys. Rev. Lett. {\bf 96}, 145701 (2006).

\bibitem{threestate2} K. Wood, C. Van den Broeck, R. Kawai, and K. Lindenberg, 
Phys. Rev. E {\bf 74}, 031113 (2006).

\bibitem{risler2} T. Risler, J. Prost, F. J\:ulicher. Phys. Rev. Lett.
{\bf 93}, 175702 (2004).

\bibitem{risler} T. Risler, J. Prost, and F. J\:ulicher, Phys. Rev. E
{\bf 72}, 016130 (2005).

\bibitem{kuznetsov} Yu. A. Kuznetsov, {\em Elements of Applied Bifurcation Theory}, 2nd ed. (Springer, New York, 1998).

\bibitem{acebron} Acebron, J., L. Bonilla, C. Perez Vicente,
F. Ritort, and R Spigler.  Rev. Mod. Phys. {\bf 77}, (2005).

\bibitem{grigolini}
S. Bianco, E. Geneston, P. Grigolini, and M. Ignaccolo,
arXiv:cond-mat/0611035 (2006).


\end{thebibliography}
\end{document}